\documentclass[preprint,showpacs,amsmath,amssymb]{revtex4}
\usepackage{graphicx}
\usepackage{color}

\begin{document}

\title{Analytic Optimization of a MERA network and its Relevance to Quantum Integrability and Wavelet}
\author{Hiroaki Matsueda\thanks{matsueda@sendai-nct.ac.jp}}
\affiliation{
Sendai National College of Technology, Sendai 989-3128, Japan
}
\date{\today}
\begin{abstract}
I present an example of how to analytically optimize a multiscale entanglement renormalization ansatz for a finite antiferromagnetic Heisenberg chain. For this purpose, a quantum-circuit representation is taken into account, and we construct the exactly entangled ground state so that a trivial IR state is modified sequentially by operating separated entangler layers (monodromy operators) at each scale. The circuit representation allows us to make a simple understanding of close relationship between the entanglement renormalization and quantum integrability. We find that the entangler should match with the $R$-matrix, not a simple unitary, and also find that the optimization leads to the mapping between the Bethe roots and the Daubechies wavelet coefficients.
\end{abstract}
\maketitle

\section{Introduction}

Recently, the powerfulness of quantum-circuit representation and design of hierarchical tensor networks with the help of wavelet transformation is presented in the field of tensor network variational methods in condensed matter and statistical physics~\cite{XLQ,CHLXLQ,EvenblyWhite1,EvenblyWhite2,Ferris1,Ferris2}. Although the circuit representation itself has been settled long time ago in quantum algorithm theory~\cite{Deutsch1,Deutsch2,Grover}, it starts to merge with PEPS and MERA quite recently~\cite{Verstraete,Vidal}. One approach is called as exact holographic mapping (EHM) based on the Haal wavelet (or equivalently the Hadamard gate)~\cite{XLQ,CHLXLQ}, and an another approach is the modified Daubechies D4 wavelet representation of multiscale entanglement renormalization ansatz (MERA) for one-dimensional (1D) free fermions~\cite{EvenblyWhite1,EvenblyWhite2}. The circuit representation provides us with an efficient way of how to operationally introduce quantum entanglement into a trivial IR state by the exact unitary mapping, and the wavelet transform is quite essential to represent both of entanglement and renormalization group (RG) flow in the same building block. Along these lines, one motivation of this study is to provide a very simple but nice toy model to make the exact optimization of the MERA wavefunction possible for the 1D Heisenberg chain. We will find that such construction of the analytically optimized wavefunction contains enough information of the algebraic Bethe ansatz~\cite{Jimbo,Korepin,Gomez}. Furthermore, the two-string solution of the Bethe equation is directly related to the rotation angle of the unitary circuit that realizes the Daubechies D4 wavelet. Thus the result manifests efficiency of the previous wavelet works~\cite{EvenblyWhite1,EvenblyWhite2}.

Before going into technical details, it is helpful for reades to mention the classification of the tensor network algorithms. There are two main classes of tensor-network-type variational ansatz, depending on criticality of our target model and the corresponding entanglement-entropy scaling. One is so-called projected entangled-pair state (PEPS) class. This class can safely represent the gapped quantum systems, and the matrix product state (MPS) is a member of this class in 1D cases. The other one is called MERA class, and this is applied to critical systems. It is possible to apply PEPS to critical cases, but we need to take enormously large truncation number. Thus it is not practical. Those classifications are very nice in numerical simulations, but at the same time some ambiguities originating in the practicalness still exist. At least theoretically, MERA and PEPS seem to be everytime convertible with each other and their difference is just owing to numerical efficiency, if we can take large enough tensor dimension of PEPS. According to the exactly solvable models, it has actually been known that the Heisenberg model, which is a typical critical system, can be exactly solved by the Bethe ansatz. However, the algebraic Bethe ansatz is equivalently mapped onto MPS (matrix product Bethe ansatz)~\cite{Alcaraz1,Alcaraz2,Alcaraz3,Katsura,Maruyama}, and thus is a member of the PEPS class, not the critical MERA class. Furthermore, it is easy to show graphical representation of the mapping~\cite{EvenblyVidal1,EvenblyVidal2,EvenblyVidal3}. Therefore, the point for better understanding is to make it clear mathematically to what kinds of circuit pieces in MERA have important roles as same as that of the Yang-Baxter equation or the Bethe equation in the Bethe ansatz. The MERA contains entangler layers, and this form is quite similar to the $R$-matrix or the Lax operator in the algebraic Bethe ansatz. This would be a good hint to resolve this problem. The Bethe equation is a kind of consistency conditions for constructing the exact wavefunction. Thus, we naturally expect that the consistency would be converted to the optimization condition for the MERA trial function. We will focus this point later.

The organization of this paper is as follows. In the next section, we explain a method of analytical optimization of MERA for a Heisenberg chain. In Sec.III, we discuss implications for the quantum integrability as well as Daubechies wavelet contained in the circuit representation of the MERA network. Finally, we summarize this work.

\section{Quantum Circuit Represenrtation of MERA Network}

\subsection{Preliminaries}

Let us first examine some properties of the MERA network as a unitary operation circuit. For this purpose, we start with a spin-$1/2$ $4$-site antiferromagnetic Heisenberg model
\begin{eqnarray}
H=\sum_{i}\vec{S}_{i}\cdot\vec{S}_{i+1}=\vec{S}_{1}\cdot\vec{S}_{2}+\vec{S}_{2}\cdot\vec{S}_{3}+\vec{S}_{3}\cdot\vec{S}_{4}\left(+\vec{S}_{4}\cdot\vec{S}_{1}\right),
\end{eqnarray}
where the last term is introduced in the periodic boundary condition. The Hilbert space of this model is spanned by ${\cal H}=V^{\otimes 4}$ with the computational basis (a two-dimensional complex vector space) $V=\mathbb{C}^{2}$. We denote $\left|0\right>=\left|\uparrow\right>=(1,0)^{t}$ and $\left|1\right>=\left|\downarrow\right>=(0,1)^{t}$. This model is of course exactly solvable by directly diagonalizing the Hamiltonian matrix. By using this simple model, we would like to examine the functionality and algebraic properties of each tensor of the MERA network.

In the open boundary case, the variational wavefunction of MERA is given by
\begin{eqnarray}
\left|\psi\right>=\sum_{s_{1},s_{2},s_{3},s_{4}}\sum_{\alpha,\beta}\sum_{\gamma,\delta}T^{\gamma\delta}L_{\gamma}^{s_{1}\alpha}R_{\delta}^{s_{4}\beta}U_{\alpha\beta}^{s_{2}s_{3}}\left|s_{1}s_{2}s_{3}s_{4}\right>,
\end{eqnarray}
where $T$ is the top tensor, $L,R$ are the isometries, and $U$ is the entangler tensor. Since we consider the unitary mapping, the indices $\alpha$ and $\beta$ take $0$ or $1$. When we do not take coarse graining, the indices $\gamma$ and $\delta$ run from $0$ to $3$. In general, all the indices are truncated up to $\chi$ degrees of freedom in the case of scale-invariant MERA. Taking the 4-site model is minimal requirement for introducing the entangler tensor. We transform the above trial function into the following form:
\begin{eqnarray}
\left|\psi\right>=\sum_{\alpha,\beta}\sum_{\gamma,\delta}T^{\gamma\delta}\left(\sum_{s_{1}}L_{\gamma}^{s_{1}\alpha}\left|s_{1}\right>\right)\otimes\left(\sum_{s_{2},s_{3}}U_{\alpha\beta}^{s_{2}s_{3}}\left|s_{2}s_{3}\right>\right)\otimes\left(\sum_{s_{4}}R_{\delta}^{s_{4}\beta}\left|s_{4}\right>\right),
\end{eqnarray}
and we regard this as the inverse RG flow from the IR limit to UV:
\begin{eqnarray}
T \Rightarrow \left(L\otimes R\right)T \Rightarrow \left(\mathbb{I}_{2}\otimes U(\theta)\otimes\mathbb{I}_{2}\right)\left(L\otimes R\right)T,
\end{eqnarray}
where $\mathbb{I}_{d}$ is the $d\times d$ unit matrix. Namely, we start with a trivial (classical, weak entanglement) state $\left|T\right>$, and gradually introduce stronger entanglement by the entangler $U$.

Now we consider a small finite lattice. Thus there is an energy gap according to the finite-size cut-off, even though we would like to finally understand fixed-point behavior of RG. Therefore, we expect that the truncation of upper MERA layers do not affect severely to the exact representation of the wave function. Therefore, we approximate
\begin{eqnarray}
\left|\psi\right>\simeq\sum_{\alpha,\beta}\left(\sum_{s_{1}}L^{s_{1}\alpha}\left|s_{1}\right>\right)\otimes\left(\sum_{s_{2},s_{3}}U_{\alpha\beta}^{s_{2}s_{3}}\left|s_{2}s_{3}\right>\right)\otimes\left(\sum_{s_{4}}R^{s_{4}\beta}\left|s_{4}\right>\right),
\end{eqnarray}
except for a normalization factor. When we apply the singular value decomposition (SVD) to $U$ as
\begin{eqnarray}
U_{\alpha\beta}^{s_{2}s_{3}}\rightarrow U_{(s_{2}\alpha)(s_{3}\beta)}=\sum_{l=1}^{\chi}A_{l}(s_{2}\alpha)\sqrt{\Lambda_{l}}B_{l}(s_{3}\beta),
\end{eqnarray}
and this is practically equal to MPS. Actually, we obtain
\begin{eqnarray}
\left|\psi\right>\simeq\sum_{l=1}^{\chi}\sqrt{\Lambda_{l}}\left(\sum_{\alpha}\sum_{s_{1},s_{2}}L^{s_{1}\alpha}A_{l}(s_{2}\alpha)\left|s_{1}s_{2}\right>\right)\otimes\left(\sum_{\beta}\sum_{s_{3},s_{4}}R^{s_{4}\beta}B_{l}(s_{3}\beta)\left|s_{3}s_{4}\right>\right).
\end{eqnarray}
Here, the matrix dimension is equal to $\chi=4$, and it is basically possible to exactly optimize this wavefunction. When we regard this as MPS, $L$, $R$, and $U$ are equally treated. On the other hand, in the present case, we first apply $L$ and $R$ to $T$, and then apply $U$ to it. Thus we introduce the order of operation of each layer. At the same time, each tensor has clear physical meaning: for instance $U$ is the unitary operator and $L$ and $R$ are isometries (or coarse-graining operations). Therefore, '{\it order of operations}' and '{\it functionality of each tensor}' are two important properties of circuit representation and design.

\begin{figure}[htbp]
\begin{center}
\includegraphics[width=12cm]{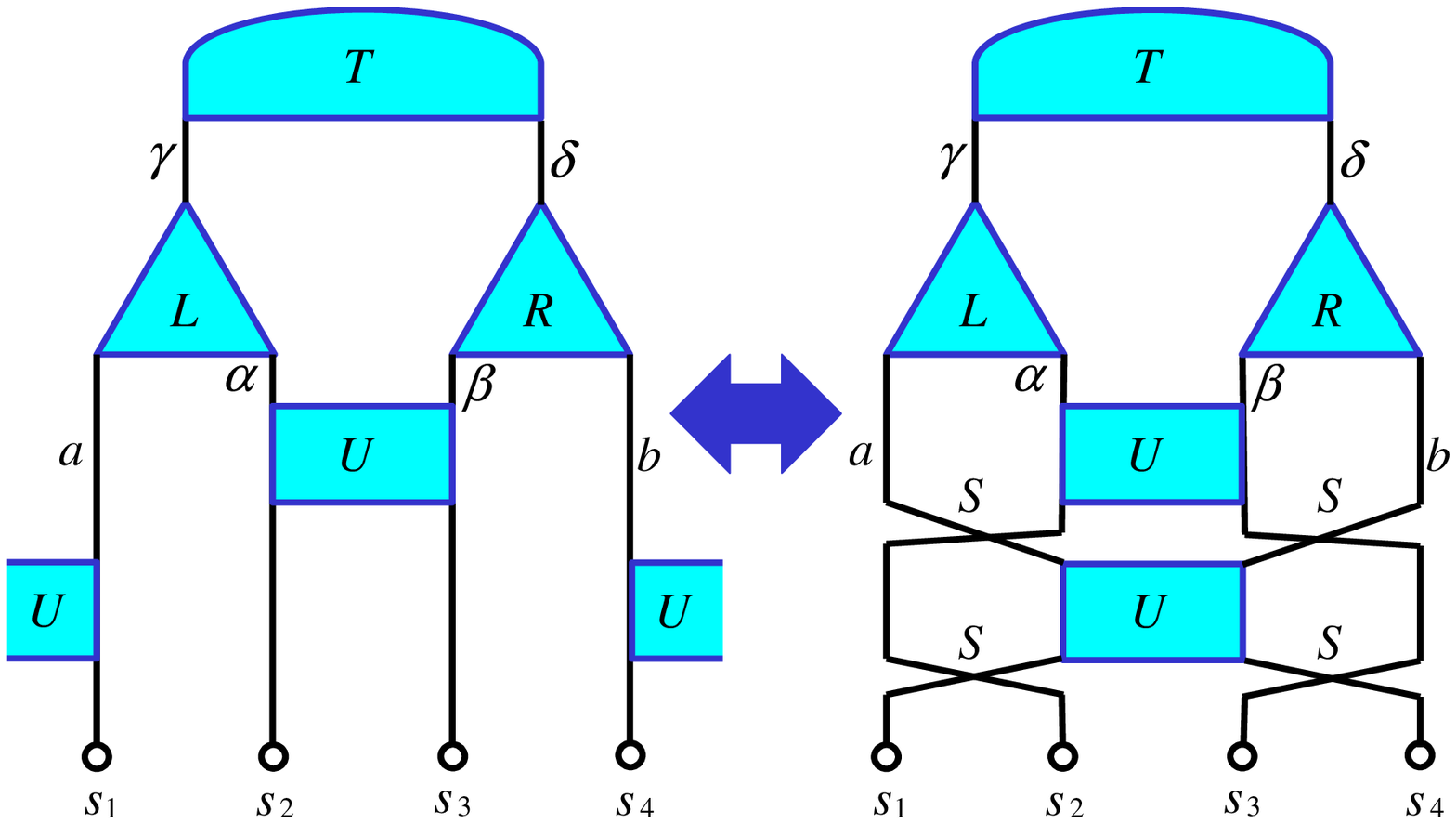}
\end{center}
\caption{Unitary circuit representation of the MERA network.}
\label{snct70fig1}
\end{figure}

\begin{figure}[htbp]
\begin{center}
\includegraphics[width=12cm]{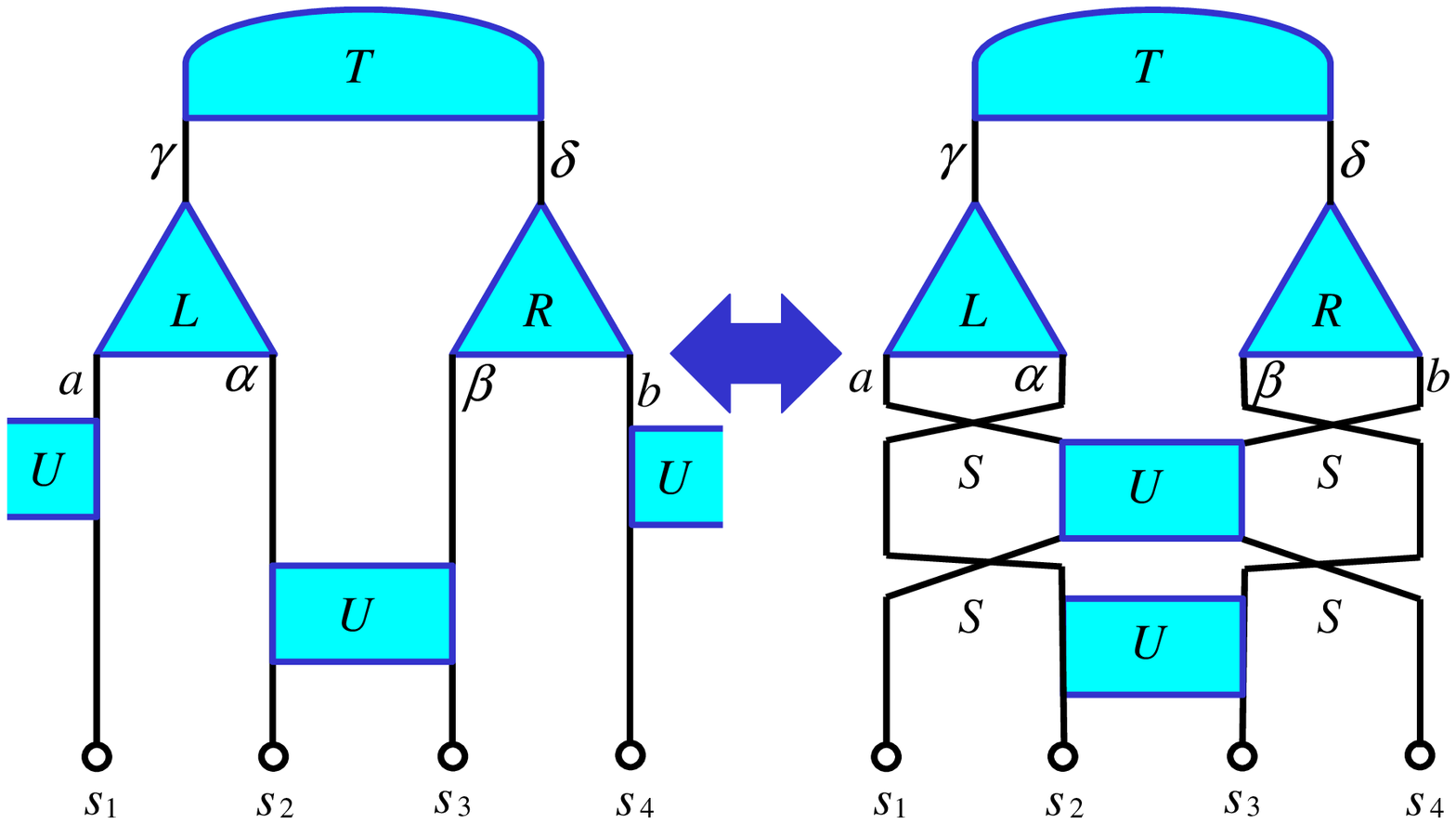}
\end{center}
\caption{Exchange symmetry of two entanglers}
\label{snct70fig2}
\end{figure}

In the following, we mainly focus on the periodic boundary condition. For the periodic boundary case, $\left|s_{1}\right>$ and $\left|s_{4}\right>$ are also intertwinned by the entangler, and then the form of the wavefunction is given by
\begin{eqnarray}
\left|\psi\right>=\sum_{s_{1},s_{2},s_{3},s_{4}}\sum_{\alpha,\beta}\sum_{\gamma,\delta}\sum_{a,b}T^{\gamma\delta}U_{ab}^{s_{4}s_{1}}L_{\gamma}^{a\alpha}R_{\delta}^{b\beta}U_{\alpha\beta}^{s_{2}s_{3}}\left|s_{1}s_{2}s_{3}s_{4}\right>.
\end{eqnarray}
In this case, we need some effort to introduce the hierarchical representation of the network. The key ingredient is the swap gate or the permutation operator $S$ defined by
\begin{eqnarray}
S=\left(\begin{array}{cccc}1&0&0&0\\0&0&1&0\\0&1&0&0\\0&0&0&1\end{array}\right) \; , \; S\left(\left|s\right>\otimes\left|s^{\prime}\right>\right)=\left|s^{\prime}\right>\otimes\left|s\right>.
\end{eqnarray}
Since $S^{2}=\mathbb{I}_{4}$, $S$ is unitary operation :$S=S^{\dagger}$. The trial function can then be represented as
\begin{eqnarray}
\left|\psi\right>=\left(S\otimes S\right)\left(\mathbb{I}_{2}\otimes U(\theta)\otimes\mathbb{I}_{2}\right)\left(S\otimes S\right)\left(\mathbb{I}_{2}\otimes U(\theta)\otimes\mathbb{I}_{2}\right)\left|\Omega\right>,
\end{eqnarray}
where the fixed point is represented as
\begin{eqnarray}
\left|\Omega\right>=(L\otimes R)\left|T\right>.
\end{eqnarray}
We assume that each unitary gate has one variational parameter $\theta$, and that two unitary gates are equivalent with each other. We think the structure of scale-invariant MERA network. The graphical representation of this state is given in Fig.~\ref{snct70fig1}. When we define $(S\otimes S)\left|\psi\right>=\left|\psi^{\prime}\right>$, it is worth mentioning that the following transformation exists
\begin{eqnarray}
\left|\psi^{\prime}\right>=\left(\mathbb{I}_{2}\otimes U(\theta)\otimes\mathbb{I}_{2}\right)\left(S\otimes S\right)\left(\mathbb{I}_{2}\otimes U(\theta)\otimes\mathbb{I}_{2}\right)\left(S\otimes S\right)\left|\Omega^{\prime}\right>.
\end{eqnarray}
This symmetry originates in the invariance of exchange between two operations $\mathbb{I}_{2}\otimes U(\theta)\otimes\mathbb{I}_{2}$ and $\left(S\otimes S\right)\left(\mathbb{I}_{2}\otimes U(\theta)\otimes\mathbb{I}_{2}\right)\left(S\otimes S\right)$ (see Fig.~\ref{snct70fig2}):
\begin{eqnarray}
\left[ \mathbb{I}_{2}\otimes U(\theta)\otimes\mathbb{I}_{2} , \left(S\otimes S\right)\left(\mathbb{I}_{2}\otimes U(\theta)\otimes\mathbb{I}_{2}\right)\left(S\otimes S\right) \right]=0.
\end{eqnarray}
This equality can be proved by the direct calculation of the matrix elements. This algebra is essentially equal to that for monodromy and transfer matrices in the algebraic Bethe ansatz. It is easy to generalize this result to more complicated cases, since there is not direct overlap among entanglers.

\subsection{Control of Entanglement by Unitary Operation}

Let us explain how to optimize the abovementioned trial function. For this purpose, we assume that in the top layer the indices $\gamma$ and $\delta$ are irrelevant after RG, and we take
\begin{eqnarray}
\left|\Omega\right>=(L\otimes R)\left|T\right>=\left(\begin{array}{c}L^{00}\\ L^{01}\\ L^{10}\\ L^{11}\end{array}\right)\otimes\left(\begin{array}{c}R^{00}\\ R^{01}\\ R^{10}\\ R^{11}\end{array}\right),
\end{eqnarray}
where we have omitted the degrees of freedom of $\gamma$ and $\delta$ and then $T$ is just a scalar variable (omitted here). The isometry conditions are represented as
\begin{eqnarray}
(L^{00})^{2}+(L^{01})^{2}+(L^{10})^{2}+(L^{11})^{2}=1,
\end{eqnarray}
and
\begin{eqnarray}
(R^{00})^{2}+(R^{01})^{2}+(R^{10})^{2}+(R^{11})^{2}=1.
\end{eqnarray}

Let us next consider the operator $U(\theta)$ which is unitary $U(\theta)U^{\dagger}(\theta)=\mathbb{I}_{4}$. This operator can be factorized as
\begin{eqnarray}
U(\theta)=\left(\begin{array}{cccc}1&0&0&0\\0&\cos\theta&\sin\theta&0\\0&-\sin\theta&\cos\theta&0\\0&0&0&1\end{array}\right),
\end{eqnarray}
where we only consider unitary evolution processes keeping the spin quantum number. The matrix $U(\theta)$ except for the minus sign is almost equal to the $R$-matrix in the algebraic Bethe ansatz. The index $\theta$ is thus corresponding to the spectral parameter. The operation of unitary layer at a particular scale corresponds to generating a monodromy matrix. We obtain
\begin{eqnarray}
\mathbb{I}_{2}\otimes U(\theta)\otimes\mathbb{I}_{2}=\left(\begin{array}{cc}1&0\\ 0&1\end{array}\right)\otimes\left(\begin{array}{cccc}1&0&0&0\\0&\cos\theta&\sin\theta&0\\0&-\sin\theta&\cos\theta&0\\0&0&0&1\end{array}\right)\otimes\left(\begin{array}{cc}1&0\\ 0&1\end{array}\right)=\left(\begin{array}{cc}R_{8}&0\\ 0&R_{8}\end{array}\right),
\end{eqnarray}
and
\begin{eqnarray}
R_{8}=\left(\begin{array}{cccc}1&0&0&0\\0&\cos\theta&\sin\theta&0\\0&-\sin\theta&\cos\theta&0\\0&0&0&1\end{array}\right)\otimes\left(\begin{array}{cc}1&0\\ 0&1\end{array}\right)=\left(\begin{array}{cccccccc}1&0&0&0&0&0&0&0\\0&1&0&0&0&0&0&0\\0&0&\cos\theta&0&\sin\theta&0&0&0\\0&0&0&\cos\theta&0&\sin\theta&0&0\\0&0&-\sin\theta&0&\cos\theta&0&0&0\\0&0&0&-\sin\theta&0&\cos\theta&0&0\\0&0&0&0&0&0&1&0\\0&0&0&0&0&0&0&1\end{array}\right).
\end{eqnarray}
We will later regard the spectral parameter $\theta$ as a variational parameter so that $\left|\psi\right>$ becomes the exact ground state.

Now the trial wavefunction is represented as
\begin{eqnarray}
\left|\psi\right>=\left(S\otimes S\right)\left(\begin{array}{cc}R_{8}&\\&R_{8}\end{array}\right)\left(S\otimes S\right)\left(\begin{array}{cc}R_{8}&\\&R_{8}\end{array}\right)\left(\begin{array}{c}\left|W_{L}\right>\\ \left|W_{R}\right>\end{array}\right),
\end{eqnarray}
where $\left|W_{L}\right>,\left|W_{R}\right>$ are defined respectively as
\begin{eqnarray}
\left|W_{L}\right>=\left(\begin{array}{c}L^{00}R^{00}\\L^{00}R^{01}\\L^{00}R^{10}\\L^{00}R^{11}\\L^{01}R^{00}\\L^{01}R^{01}\\L^{01}R^{10}\\L^{01}R^{11}\end{array}\right) \; , \; \left|W_{R}\right>=\left(\begin{array}{c}L^{10}R^{00}\\L^{10}R^{01}\\L^{10}R^{10}\\L^{10}R^{11}\\L^{11}R^{00}\\L^{11}R^{01}\\L^{11}R^{10}\\L^{11}R^{11}\end{array}\right).
\end{eqnarray}
Here $S\otimes S$ is represented as
\begin{eqnarray}
S\otimes S=\left(\begin{array}{cccc}S&0&0&0\\0&0&S&0\\0&S&0&0\\0&0&0&S\end{array}\right),
\end{eqnarray}
and we obtain
\begin{eqnarray}
\left(\begin{array}{cccc}S&0&0&0\\0&0&S&0\\0&S&0&0\\0&0&0&S\end{array}\right)\left(\begin{array}{cc}R_{8}&0\\0&R_{8}\end{array}\right)=\left(\begin{array}{cc}\left(\begin{array}{cc}S&0\\0&0\end{array}\right)R_{8}&\left(\begin{array}{cc}0&0\\S&0\end{array}\right)R_{8}\\ \left(\begin{array}{cc}0&S\\0&0\end{array}\right)R_{8}&\left(\begin{array}{cc}0&0\\0&S\end{array}\right)R_{8}\end{array}\right).
\end{eqnarray}
The trial wavefunction is then given by
\begin{eqnarray}
\left|\psi\right>=\left(\begin{array}{c}\left|\psi_{L}\right>\\ \left|\psi_{R}\right>\end{array}\right)=\left(\begin{array}{cc}\left(\begin{array}{cc}S&0\\0&0\end{array}\right)R_{8}&\left(\begin{array}{cc}0&0\\S&0\end{array}\right)R_{8}\\ \left(\begin{array}{cc}0&S\\0&0\end{array}\right)R_{8}&\left(\begin{array}{cc}0&0\\0&S\end{array}\right)R_{8}\end{array}\right)^{2}\left(\begin{array}{c}\left|W_{L}\right>\\ \left|W_{R}\right>\end{array}\right).
\end{eqnarray}
Hereafter we just consider $\left|\psi_{L}\right>$ owing to the symmetry between $\left|\psi_{L}\right>$ and $\left|\psi_{R}\right>$. The $8$-dimensional state $\left|\psi_{L}\right>$ can be evaluated as
\begin{eqnarray}
\left|\psi_{L}\right> &=&\left[\left(\begin{array}{cc}S&0\\0&0\end{array}\right)R_{8}\left(\begin{array}{cc}S&0\\0&0\end{array}\right)R_{8}+\left(\begin{array}{cc}0&0\\S&0\end{array}\right)R_{8}\left(\begin{array}{cc}0&S\\0&0\end{array}\right)R_{8}\right]\left|W_{L}\right> \nonumber \\
&& +\left[\left(\begin{array}{cc}S&0\\0&0\end{array}\right)R_{8}\left(\begin{array}{cc}0&0\\S&0\end{array}\right)R_{8}+\left(\begin{array}{cc}0&0\\S&0\end{array}\right)R_{8}\left(\begin{array}{cc}0&0\\0&S\end{array}\right)R_{8}\right]\left|W_{R}\right>.
\end{eqnarray}
The explict forms of these matrix elements are
\begin{eqnarray}
\left|\psi_{L}\right>=\left(\begin{array}{cccccccc}1&0&0&0&0&0&0&0\\0&c&0&0&0&0&0&0\\0&0&c&0&s&0&0&0\\0&0&0&c^{2}&0&cs&0&0\\0&0&-s&0&c&0&0&0\\0&0&0&-cs&0&c^{2}&0&0\\0&0&0&0&0&0&1&0\\0&0&0&0&0&0&0&c\end{array}\right)\left|W_{L}\right>+\left(\begin{array}{cccccccc}0&0&0&0&0&0&0&0\\s&0&0&0&0&0&0&0\\0&0&0&0&0&0&0&0\\0&0&cs&0&s^{2}&0&0&0\\0&0&0&0&0&0&0&0\\0&0&-s^{2}&0&cs&0&0&0\\0&0&0&0&0&0&0&0\\0&0&0&0&0&0&s&0\end{array}\right)\left|W_{R}\right>,
\end{eqnarray}
where we use the abbreviations $c=\cos\theta,s=\sin\theta$. Clearly, the bases $0011$ and $0101$ hybridize with each other, and the hybridization induces entanglement. This procedure corresponds to making a singlet by the entangler.

According to the symmetry in the spin space (exchangeability between $0$ and $1$), we find
\begin{eqnarray}
(W_{L})_{h}=(W_{R})_{8-h},
\end{eqnarray}
for $h=1,2,...,8$, and exchange the order of the vector elements by using this relation. Then we obtain
\begin{eqnarray}
\left|\psi_{L}\right>=\left(\begin{array}{cccccccc}1&0&0&0&0&0&0&0\\0&c&0&0&0&0&0&s\\0&0&c&0&s&0&0&0\\0&0&0&1&0&2cs&0&0\\0&0&-s&0&c&0&0&0\\0&0&0&0&0&c^{2}-s^{2}&0&0\\0&0&0&0&0&0&1&0\\0&s&0&0&0&0&0&c\end{array}\right)\left(\begin{array}{c}L^{00}R^{00}\\L^{00}R^{01}\\L^{00}R^{10}\\L^{00}R^{11}\\L^{01}R^{00}\\L^{01}R^{01}\\L^{01}R^{10}\\L^{01}R^{11}\end{array}\right).
\end{eqnarray}

Here, we define a trivial state by introducing
\begin{eqnarray}
L^{00}=R^{00}=R^{11}=0,
\end{eqnarray}
as a weaker entangled state (more precisely speaking they are high energy states with partially ferromagnetic configuration), and then we obtain
\begin{eqnarray}
\left|\psi_{L}\right>=\left(\begin{array}{cccccccc}1&0&0&0&0&0&0&0\\0&c&0&0&0&0&0&s\\0&0&c&0&s&0&0&0\\0&0&0&1&0&2cs&0&0\\0&0&-s&0&c&0&0&0\\0&0&0&0&0&c^{2}-s^{2}&0&0\\0&0&0&0&0&0&1&0\\0&s&0&0&0&0&0&c\end{array}\right)\left(\begin{array}{c}0\\0\\0\\0\\0\\L^{01}R^{01}\\L^{01}R^{10}\\0\end{array}\right)=\left(\begin{array}{c}0\\0\\0\\2csL^{01}R^{01}\\0\\ \left(c^{2}-s^{2}\right)L^{01}R^{01}\\L^{01}R^{10}\\0\end{array}\right).
\end{eqnarray}
Clearly, the entanglement entropy increases, since the basis $0011$ has finite amount of weight after quantum operation.

\subsection{Comparison with the Exact Diagonalization Result}

In the previous subsection, we regard $\theta$ as a variational parameter. In order to examine how precise the abovementioned ansatz is, we optimize it for the 4-site antiferromagnetic Heisenberg model by varying the magnitude of $\theta$. The Hamiltonian matrix for the Heisenberg model under the periodic boundary condition is given by
\begin{eqnarray}
H=\left(\begin{array}{cccccc}0&\frac{1}{2}&0&0&\frac{1}{2}&0\\ \frac{1}{2}&-1&\frac{1}{2}&\frac{1}{2}&0&\frac{1}{2}\\0&\frac{1}{2}&0&0&\frac{1}{2}&0\\0&\frac{1}{2}&0&0&\frac{1}{2}&0\\ \frac{1}{2}&0&\frac{1}{2}&\frac{1}{2}&-1&\frac{1}{2}\\0&\frac{1}{2}&0&0&\frac{1}{2}&0\end{array}\right) \; , \; \left(\begin{array}{c} \uparrow\uparrow\downarrow\downarrow\\ \uparrow\downarrow\uparrow\downarrow\\ \uparrow\downarrow\downarrow\uparrow\\ \downarrow\uparrow\uparrow\downarrow\\ \downarrow\uparrow\downarrow\uparrow\\ \downarrow\downarrow\uparrow\uparrow\end{array}\right)\Leftrightarrow\left(\begin{array}{c}0011\\0101\\0110\\1001\\1010\\1100\end{array}\right).
\end{eqnarray}
The ground state after the exact diagonalization is given by
\begin{eqnarray}
\left|\psi\right>=A\left|\uparrow\uparrow\downarrow\downarrow\right>+B\left|\uparrow\downarrow\uparrow\downarrow\right>+C\left|\uparrow\downarrow\downarrow\uparrow\right>+D\left|\downarrow\uparrow\uparrow\downarrow\right>+E\left|\downarrow\uparrow\downarrow\uparrow\right>+F\left|\downarrow\downarrow\uparrow\uparrow\right>,
\end{eqnarray}
with coefficients
\begin{eqnarray}
A=1 \; , \; B=-2 \; , \; C=1 \; , \; D=1 \; , \; E=-2 \; , \; F=1,
\end{eqnarray}
(fininally we need to normalize it). This can be represented as the resonating valence bond state given by
\begin{eqnarray}
\left|\psi\right> &=& \left|\uparrow\uparrow\downarrow\downarrow\right>-2\left|\uparrow\downarrow\uparrow\downarrow\right>+\left|\uparrow\downarrow\downarrow\uparrow\right>+\left|\downarrow\uparrow\uparrow\downarrow\right>-2\left|\downarrow\uparrow\downarrow\uparrow\right>+\left|\downarrow\downarrow\uparrow\uparrow\right> \nonumber \\
&=& \left(\left|\uparrow\right>_{1}\otimes\left|\downarrow\right>_{4}-\left|\downarrow\right>_{1}\otimes\left|\uparrow\right>_{4}\right)\left(\left|\uparrow\right>_{2}\otimes\left|\downarrow\right>_{3}-\left|\downarrow\right>_{2}\otimes\left|\uparrow\right>_{3}\right) \nonumber \\
&& + \left(\left|\uparrow\right>_{1}\otimes\left|\downarrow\right>_{2}-\left|\downarrow\right>_{1}\otimes\left|\uparrow\right>_{2}\right)\left(\left|\uparrow\right>_{3}\otimes\left|\downarrow\right>_{3}-\left|\downarrow\right>_{3}\otimes\left|\uparrow\right>_{3}\right).
\end{eqnarray}

We find that the abovementioned variational ansatz has none-zero weights on those basis states. Actually, the correspondence is given by
\begin{eqnarray}
A=2csL^{01}R^{01} \; , \; B=\left(c^{2}-s^{2}\right)L^{01}R^{01} \; , \; C=L^{01}R^{10}.
\end{eqnarray}
Except for the overall normalization factor, we find
\begin{eqnarray}
r=-\frac{R^{01}}{R^{10}} \; , \; A=r\sin\left(-2\theta\right) \; , \; B=-r\cos\left(-2\theta\right) \; , \; C=1.
\end{eqnarray}
Finally, we obtain
\begin{eqnarray}
\sin(-2\theta)=\frac{1}{\sqrt{5}} \; , \; \cos(-2\theta)=\frac{2}{\sqrt{5}} \; , \; r=\sqrt{5},
\end{eqnarray}
and can determine the value of the variational parameter $\theta$. We numerically obtain $\theta\sim -0.074\pi$. This $\theta$ value is close to $-\pi/12\sim -0.083\pi$, indicating that this would be related to the Daubechies D4 wavelet~\cite{EvenblyWhite1,EvenblyWhite2}.

\section{Implications to the Algebraic Bethe Ansatz}

\subsection{Lax and Monodromy Matrices}

In the previous section, we have examined the small $4$-site system, but still we found rich aspects associated with quantum integrability and wavelet. Here, we generalize the previous discussion so that we understand which quantities are directly related to necessary tools in the algebraic Bethe ansatz such as the Lax operator and the monodromy matrix.

\begin{figure}[htbp]
\begin{center}
\includegraphics[width=14cm]{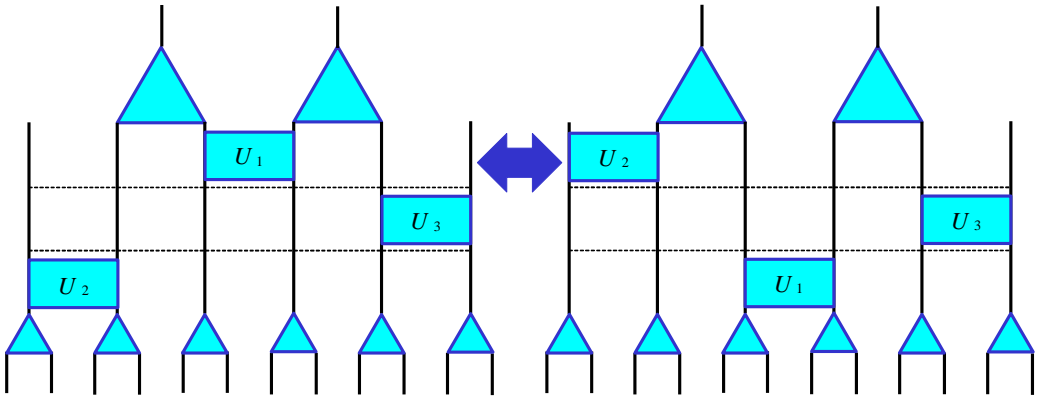}
\end{center}
\caption{Exchangeability of unitary operations at a particular length scale.}
\label{snct70fig3}
\end{figure}

At the $k$-th step of the inverse RG transformation from the IR limit $\left|T\right>$, the network has the $\left(2^{k}-1\right)$-entanglers. We denote each entangler tensor as
\begin{eqnarray}
{\cal U}_{j}(\lambda)=\underbrace{\mathbb{I}\otimes\cdots\otimes\mathbb{I}}_{j-1}\otimes U(\lambda)\otimes\underbrace{\mathbb{I}\otimes\cdots\otimes\mathbb{I}}_{2^{k}-1-j},
\end{eqnarray}
where we replace $\theta$ with $\lambda$, since we take gengral $\chi$ degrees of freedom in the auxiliary space and the unitary matrix may not be equal to simple rotation one on 2D space. Note that now we do not consider any coarse graining at the position of isometry tensors for avoiding confusion. Clearly we have the following property for ${\cal U}_{j}(\lambda)$:
\begin{eqnarray}
\left[ {\cal U}_{i}(\lambda), {\cal U}_{j}(\lambda)\right]=0, \label{UU}
\end{eqnarray}
since each entangler tensor is separately located and there is no direct overlap at the present length scale. The exchangeability means that the result does not depend on the order of product of ${\cal U}_{j}(\lambda)$ (see Fig.~\ref{snct70fig3}). This is very strong constraint that determines the algebraic structure of this MERA network. Then we obtain the monodromy matrix as a sequence of entanglers at the level $k$
\begin{eqnarray}
{\cal T}(\lambda)={\cal U}_{1}(\lambda){\cal U}_{2}(\lambda)\cdots{\cal U}_{2^{k}-1}(\lambda).
\end{eqnarray}
The operator ${\cal U}_{j}(\lambda)$ has some indices associated with the auxiliary space, and thus we think that ${\cal U}_{j}(\lambda)$ behaves as the local Lax matrix. Then, ${\cal T}(\lambda)$ corresponds to the monodromy matrix at the scale $k$. In the present MERA case, we explicitely use the auxiliary space indices, it is not necessary to distinguish the monodromy matrix from the transfer matrix. In the standard notation, we related $U(\lambda)$ with
\begin{eqnarray}
U(\lambda)\rightarrow\left(\begin{array}{cc}{\cal L}_{0}^{0}(\lambda)& {\cal L}_{1}^{0}(\lambda)\\ {\cal L}_{0}^{1}(\lambda)& {\cal L}_{1}^{1}(\lambda)\end{array}\right),
\end{eqnarray}
when we keep $\chi=2$ for all the RG processes. The Bethe eigenstate coarse-grained at the scale $k$ is given by
\begin{eqnarray}
\left|\Omega_{k}\right>={\cal T}_{k}(\lambda)\left|\Omega_{k-1}\right>,
\end{eqnarray}
where the number of effective sites is $2^{k}$.

\subsection{Bethe Equations and their Roots}

Let us consider the $R$-matrix in the algebraic Bethe ansatz:
\begin{eqnarray}
R(\lambda)=\left(\begin{array}{cccc}1&0&0&0\\0&b(\lambda)&c(\lambda)&0\\0&c(\lambda)&b(\lambda)&0\\0&0&0&1\end{array}\right) \longleftrightarrow U(\theta)=\left(\begin{array}{cccc}1&0&0&0\\0&\cos\theta&\sin\theta&0\\0&-\sin\theta&\cos\theta&0\\0&0&0&1\end{array}\right),
\end{eqnarray}
where we define
\begin{eqnarray}
b(\lambda)=\frac{2i}{\lambda+2i} \; , \; c(\lambda)=\frac{\lambda}{\lambda+2i},
\end{eqnarray}
and they satisfy the following properties
\begin{eqnarray}
b(\nu)+c(\nu)=1 \; , \; |b(\nu)|^{2}+|c(\nu)|^{2}=1.
\end{eqnarray}
Note that the $R$-matrix is somewhat different from our unitary matrix $U(\theta)$, although it is clear that both of them play crucial roles of creating entanglement by mixing $01$ and $10$ states. It is thus very important to examine the meaning of their difference. Exactly speaking, our special selection of $U(\theta)$ is not general, and in the previous section we have just taken the rotational matrix. We should be take care about what kind of unitary operation is reasonable and how this reason is related to consistency condition of quantum integrability. For this purpose, we consider the Bethe equations.

The Bethe equations for $L=4$ and $\lambda_{i}$ ($i=1,2$) are given by
\begin{eqnarray}
\left(\frac{\lambda_{1}+i}{\lambda_{1}-i}\right)^{4}=\frac{\lambda_{1}-\lambda_{2}+2i}{\lambda_{1}-\lambda_{2}-2i},
\end{eqnarray}
and
\begin{eqnarray}
\left(\frac{\lambda_{2}+i}{\lambda_{2}-i}\right)^{4}=\frac{\lambda_{2}-\lambda_{1}+2i}{\lambda_{2}-\lambda_{1}-2i}.
\end{eqnarray}
The $2$-string solutions are obtained as
\begin{eqnarray}
\lambda_{1}=-\lambda_{2}=\lambda=\tan\left(\frac{\pi}{6}\right)=\frac{1}{\sqrt{3}}.
\end{eqnarray}
We again find the characteristic angle $\phi=\pi/6\sim 2\theta$ that reminds us with the Daubechies D4 wavelet~\cite{EvenblyWhite1,EvenblyWhite2}. The relevance of the Daubechies wavelet optimization for the free fermion MERA seems to be related with quantum integralibity, in particular the mathematical structure of the root space of the Bethe equations. In the next subsection, we finally examine whether this guess is reasonable and how these features are incorporated into the optimized MERA network.

\subsection{Reparametrization of the Entangler Tensor for the Daubechies D4 Wavelet}

The magnitude of the Bethe roots obtained in the previous section suggests that the proper selection of the mathematical form of entangler operation should match with the $R$-matrix. To confirm this conjecture, we redefine
\begin{eqnarray}
\tilde{R}_{8}(\nu)=R(\nu)\otimes\left(\begin{array}{cc}1&0\\ 0&1\end{array}\right)=\left(\begin{array}{cccc}1&0&0&0\\0&b(\nu)&c(\nu)&0\\0&c(\nu)&b(\nu)&0\\0&0&0&1\end{array}\right)\otimes\left(\begin{array}{cc}1&0\\ 0&1\end{array}\right),
\end{eqnarray}
and use it instead of $R_{8}$ (in the present stage, it is not obvious whether $\nu$ is equal to $\lambda$). Fortunately, the $R$-matrix satisfies the unitary condition $R(\nu)R^{\dagger}(\nu)=\mathbb{I}_{4}$. Therefore, it is possible to do such replacement. Actually, $\bigl|\tilde{\psi}_{L}\bigr>$ replaced from $\left|\psi_{L}\right>$ is given by
\begin{eqnarray}
\bigl|\tilde{\psi}_{L}\bigr>=\left(\begin{array}{cccccccc}1&0&0&0&0&0&0&0\\0&b&0&0&0&0&0&c\\0&0&b&0&c&0&0&0\\0&0&0&b^{2}+c^{2}&0&2bc&0&0\\0&0&c&0&b&0&0&0\\0&0&0&0&0&b^{2}+c^{2}&0&0\\0&0&0&0&0&0&1&0\\0&c&0&0&0&0&0&b\end{array}\right)\left(\begin{array}{c}0\\0\\0\\0\\0\\L^{01}R^{01}\\L^{01}R^{10}\\0\end{array}\right)\propto\left(\begin{array}{c}0\\0\\0\\2bc\\0\\ b^{2}+c^{2}\\ \frac{R^{10}}{R^{01}}\\0\end{array}\right).
\end{eqnarray}
To fit this result with the exact ground state of the $4$-site Heisenberg model, we need to select $b(\nu)$ and $c(\nu)$ so that they satisfy
\begin{eqnarray}
2b(\nu)c(\nu) : \left(b^{2}(\nu)+c^{2}(\nu)\right) = 1 : -2,
\end{eqnarray}
and we find
\begin{eqnarray}
\nu=-4i\pm2\sqrt{3}.
\end{eqnarray}
This result gives us
\begin{eqnarray}
b(\nu)=\frac{-1\pm\sqrt{3}i}{4}=\frac{1}{2}e^{\pm\frac{2}{3}\pi i}=\frac{1}{2}e^{\pm\frac{1}{2}\pi i}e^{\mp\frac{1}{2}\pi i}e^{\pm\frac{2}{3}\pi i}=\frac{1}{2}e^{\pm\frac{1}{2}\pi i}e^{\pm\phi i}.
\end{eqnarray}
We find that the string solutions actually play crucial roles on optimizing the trial MERA wavefunction, if we decompose $R(\nu)$ into a more elementary $R$-matrix and an appropriate phase gate. According to Refs.~\cite{EvenblyWhite1,EvenblyWhite2}, it is necessary to do some '{\it preconditioning}' in the sense that a simple rotation is not perfect and some rotation matrices as well as phase gates should be combined with each other to show the correct ground state. The present result would reflect such situation for the quantum gate design.

\section{Summary}

In this paper, I have presented a simple example of how to obtain the exact MERA network for a short Heisenberg chain by based on quantum circuit design. The result actually agrees well with the exact diagonalization one. The point behind this work is the existence of symmetry associated with quantum integrability. Furthermore, we find the recently proposed MERA/wavelet theory is based on the special properties of the Bethe roots. To interpret the entangler as the $R$-matrix, slight modification of the network design or some transformation of the spectral parameter would be necessary. In the present case, we consider the simplest string solution, but we can think of more complicated cases. They would be related to other wavelet patterns, and this characterization is also an important future work.

Finally we briefly remark applicability of the present results to other related research fields. MERA has been attracted much attention for similarity with the AdS/CFT correspondence in string theory~\cite{Swingle}. Furthermore, the algebraic Bethe ansatz approach to the AdS/CFT correspondence is also an important topic~\cite{Beisert}. The present result would bridge these problems and bring more global view for those topics.

\section*{Acknowledgement}

HM acknowledges participants to my lecture in condensed matter physics summer school 2016 for fruitful discussion about this issue. This work was supported by JSPS Kakenhi Grant No.15K05222 and No.15H03652.

\end{document}